# X-Ray Polarization Study of Pulsar Wind Nebulae with eXTP: Simulation Results and Scientific Prospects

Kuan Liu[1], Fei Xie[1], Ming-Yu Ge[2], Wei Deng[1], and En-Wei Liang[1]
[1] Guangxi Key Laboratory for Relativistic Astrophysics, School of Physical Science and Technology, Guangxi University, Nanning 530004, People's Republic of China; xief@gxu.edu.cn
[2] Key Laboratory for Particle Astrophysics, Institute of High Energy Physics, Chinese Academy of Sciences, Beijing 100049, People's Republic of China

## Abstract

X-ray polarization observations of pulsar wind nebulae (PWNe) provide crucial insights into magnetic field structures and particle acceleration mechanisms. While the Imaging X-ray Polarimetry Explorer (IXPE) has made significant contributions to PWN studies, its limited effective area restricts observations to only the brightest sources, leaving many fainter nebulae unexplored. We evaluate the polarization capabilities of the enhanced X-ray Timing and Polarimetry mission (eXTP) for studying PWNe and establish a methodology for simulating eXTP Polarimetry Focusing Array observations using modified IXPEOBSSIM. We develop and validate a simulation framework with appropriate response functions and instrumental background models, conducting comprehensive simulations of 12 PWNe selected from the SNRcat catalogue across various evolutionary stages and brightness levels. Our simulations demonstrate that eXTP provides approximately a factor of 2 improvement in minimum detectable polarization at the 99% confidence level ($MDP_{99}$) compared to IXPE. For the brightest targets (N157B, G54.1+0.3, and Mouse), 1 Ms observations achieve $MDP_{99}$ values of 4%–5%. The area with significant polarization detection for extended sources like Vela PWN is nearly twice as large as achievable with IXPE. These enhanced capabilities will significantly expand the sample of PWNe with robust X-ray polarization measurements, enabling systematic studies of magnetic field structures, particle acceleration mechanisms, and PWN–environment interactions across different evolutionary phases.

*Unified Astronomy Thesaurus concepts:* Pulsar wind nebulae (2215); Polarimetry (1278); X-ray detectors (1815); Astronomical methods (1043)

## 1. Introduction

Pulsar wind nebulae (PWNe) are fascinating astronomical phenomena resulting from the interaction between the pulsar wind and its ambient medium. These remarkable structures provide crucial insights into high-energy astrophysical processes (O. Kargaltsev et al. 2015; E. Amato 2020). They serve as powerful probes of the evolution of compact objects and offer unique windows into understanding the energy conversion mechanisms from pulsar spin-down to particle energization and subsequent electromagnetic radiation (S. V. Bogovalov & D. V. Khangoulian 2002; Y. E. Lyubarsky 2002; L. Del Zanna et al. 2004; L. Sironi & A. Spitkovsky 2011).

The study of PWNe has considerably advanced in recent decades, revealing their complex evolutionary history and diverse morphologies. Current research indicates that PWNe evolve through four distinct phases (B. M. Gaensler & P. O. Slane 2006). In the early stage, PWNe expand freely within the supernova ejecta, forming a typically toroidal structure with jet-like features, like Crab, 3C 58, and MSH 11–54 (M. C. Weisskopf et al. 2000; B. M. Gaensler & B. J. Wallace 2003; P. Slane et al. 2004). In the second phase, they interact with the reverse shock of the surrounding supernova remnant (SNR), causing significant deformation and compression (Y. K. Ma et al. 2016; P. Slane et al. 2018). The third phase begins after they expand again within the shocked supernova ejecta, and often the PWN has already approached the edge of the SNR (E. van der Swaluw 2003; E. van der Swaluw et al. 2004). Finally, in the fourth stage, PWNe move beyond the boundary of their parent SNR and go into the interstellar medium (ISM), forming bow-shock structures (B. M. Gaensler et al. 2000; J. P. Halpern et al. 2001a; N. Klingler et al. 2018). These evolutionary stages are accompanied by changes in morphological structure, magnetic field configurations, and emission characteristics across the whole electromagnetic spectrum.

X-ray polarization studies provide a valuable perspective on PWNe, offering direct information about the magnetic field structure and particle acceleration mechanisms within these systems (S. P. Reynolds et al. 2012; N. Bucciantini et al. 2017; E. de Ona Wilhelmi 2017). Imaging X-ray Polarimetry Explorer (IXPE), which was launched in 2021 December and funded by the Agenzia Spaziale Italiana and National Aeronautics and Space Administration (E. Costa et al. 2001; L. Baldini et al. 2021; M. C. Weisskopf et al. 2022), marks a significant improvement in the field of X-ray polarization observation. It consists of three identical telescopes, each having a polarimeter that is based on the Gas Pixel Detector (GPD) and placed at the focus of the coaligned Wolter-1 mirror module assemblies (MMAs). The IXPE MMAs have an angular resolution in half power diameter (HPD) of ∼30″ and a field of view (FoV) of 12′.9 × 12′.9.

IXPE has made substantial contributions to PWNe research (N. Bucciantini et al. 2024). The X-ray polarization of the Crab, Vela, MSH 15-52, B0540–69, 3C 58, and G21.5-0.9 has been reported (F. Xie et al. 2022; N. Bucciantini et al. 2023, 2025; K. Liu et al. 2023; R. W. Romani et al. 2023; J. Wong et al. 2024; F. Xie et al. 2024; N. Di Lalla et al. 2025).

The observation of the Crab Nebula revealed a toroidal magnetic field structure with local polarization degree (PD) reaching up to 40%–50%. The X-ray polarization detection of the Vela PWN showed toroidal magnetic field patterns with a high integrated PD of 44% and local PD of ∼70%, approaching the theoretical upper limit allowed by synchrotron theory. In MSH 15−52, IXPE detected magnetic fields along its filamentary structures and a high-significance polarization >70% in the jet end. The observation of PWN B0540–69 shows an integrated PD of ∼20% similar to the Crab Nebula in 4–6 keV. These polarization measurements have revolutionized our understanding of PWN magnetic field geometries and particle transport processes.

Despite the significant advancements made by IXPE, several limitations continue to constrain our understanding of PWNe. Current observations are primarily restricted to the brightest PWNe due to sensitivity limitations, leaving many fainter nebulae unexplored. This observational gap limits our ability to conduct systematic studies across different evolutionary phases. Furthermore, deeper observations of individual sources would allow for more detailed time-resolved studies, potentially revealing time-dependent polarization signatures that constrain the evolution of nebular magnetic field (C. Zuo et al. 2025) and coevolution between the pulsar and PWN (M. Y. Ge et al. 2019). These observational challenges underscore the need for next-generation X-ray polarimetry missions with substantially enhanced capabilities.

The enhanced X-ray Timing and Polarimetry mission (eXTP) is designed to address these challenges (S. N. Zhang et al. 2016; S. Zhang et al. 2019; S.-N. Zhang et al. 2025). eXTP, scheduled for launch in 2030, comprises three primary scientific payloads: the Spectroscopic Focusing Array (SFA), the Polarimetry Focusing Array (PFA), and the Wide-field and Wide-band Camera (W2C). The SFA consists of six Wolter-I grazing-incidence X-ray telescopes designed for timing, spectral, and imaging observations in the energy range of 0.5–10 keV. The PFA comprises three identical telescopes optimized for X-ray imaging polarimetry, sensitive in the 2–8 keV range. The optics of PFA is similar to SFA, which is also used for IXPE polarimetric mission (B. D. Ramsey et al. 2022; L. Li et al. 2025), and the expected HPD and FoV of PFA are 30″ and 9′.8 × 9′.8. The W2C is designed for discovering new high-energy bursts and triggering rapid onboard autonomous follow-up observations of the SFA and PFA, with an FoV of 1500 square degrees and an energy range of 30–600 keV.

The PFA detectors consist of three key components: calibration and filter wheel, detectors, and back-end electronics. The calibration sources will be used to calibrate or monitor the performance of the detector throughout the mission lifetime, including gain, energy resolution, modulation factor, and spurious modulation. The gray filter is intended for observations of very bright sources that may produce high count rates and lead to severe pileup. The GPD is adopted as the focal plane polarimeter for the PFA. It measures the 2D ionization track of the photoelectron in the gas chamber and infers the polarization of the X-ray source by reconstructing the emission angle of the photoelectron.

The PFA and IXPE are structurally similar in general. The effective area of the PFA is about 5 times that of IXPE at 3 keV. In the case of negligible background, an exposure of 1 ks of the Crab Nebula will result in a minimum detectable polarization (MDP) of 2%, and the mean modulation factor in 2–8 keV is ∼0.23 weighted by the observed spectrum.

Simulations are indispensable and ubiquitous tools for the development and optimization of modern astronomical telescopes. L. Qi et al. (2022) performed the observation simulation of PFA with GEANT4 toolkit (S. Agostinelli et al. 2003). Based on their works, we propose a more lightweight approach for rapid simulation of PFA observations and employ this methodology to explore the PWN physics in the eXTP era. This paper is organized as follows: Section 2 describes our simulation methodology; Section 3 validates our simulation approach; Section 4 presents the simulation results for 12 PWNe and demonstrates eXTP's enhanced capabilities; Section 5 discusses the astrophysical implications; and Section 6 concludes with a summary of our findings and future prospects.

## 2. Prescription for eXTP Simulation

Our approach adopts the IXPEOBSSIM framework. IXPEOBSSIM is a software package specifically developed for the IXPE mission (L. Baldini et al. 2022). The eXTP-PFA polarimeter employs a GPD that is analogous to IXPE. This fundamental similarity allows us to utilize the IXPEOBSSIM framework for eXTP-PFA studies with appropriate modifications.

The XPOBSSIM and XPBIN tools of IXPEOBSSIM are designed for simulating and analyzing IXPE observations.[3] XPOBSSIM takes a detailed source model along with parameterized response files to produce an event list similar to the data format of a real observation. The source model comprises spectral, temporal, morphological, and polarization attributes, while the response files include the effective area function, point-spread function (PSF), redistribution matrix function, and modulation factor function. After simulating the observation with XPOBSSIM, XPBIN will process the event lists and generate binned data products, including Stokes spectra ($I$, $Q$, and $U$) and various data structures that support analysis. The complete simulation framework is illustrated in Figure 1.

To simulate observations for the eXTP-PFA, it is necessary to substitute the response function files of XPOBSSIM with those of the eXTP-PFA. In addition, the ebounds.py file located in the ∼IXPEOBSSIM/IRF directory plays a crucial role in supplying the relationship between the channel and energy, and the corresponding values must be updated. Furthermore, the eXTP-PFA has a shorter dead time ∼0.28 ms compared to IXPE, leading to a dead time correction rate of 15% for the Crab complex (a fluence of $2 \times 10^{-8}$ erg cm$^{-2}$ s$^{-1}$ in the 2–8 keV range). This could be easily handled by setting "deadtime = 0.00028" in XPOBSSIM.

Instrumental background is vital for analyzing faint sources, particularly for the dim and extended PWNe. To achieve this, we use a phenomenological model comprising three power-law components POWERLAW+POWERLAW+POWERLAW, as depicted in Figure 2, to model the background spectrum of eXTP-PFA, which is obtained from GEANT4 simulation. The fitting results are detailed in Table 1. The model well reproduces the background spectrum, and the residual uniformly distributes around the model line, which demonstrates that this model provides a good representation of the instrumental background.

---

[3] IXPEOBSSIM documentation: https://ixpeobssim.readthedocs.io/en/latest/?badge=latest.

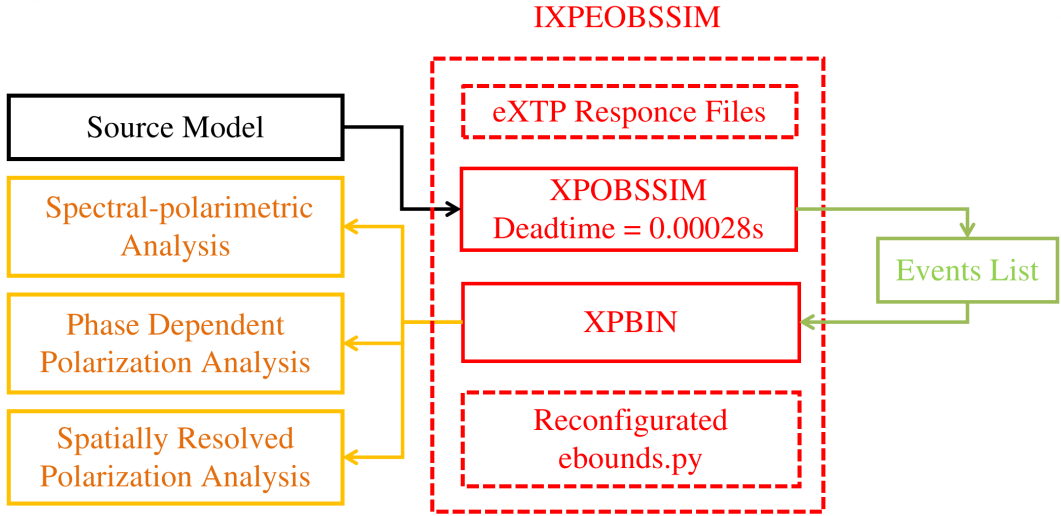

**Figure 1.** Flowchart of the eXTP-PFA simulation methodology. The diagram illustrates the complete simulation framework from input models (eXTP Response Files and Source Model) through the modified IXPEOBSSIM pipeline to the final analysis products.

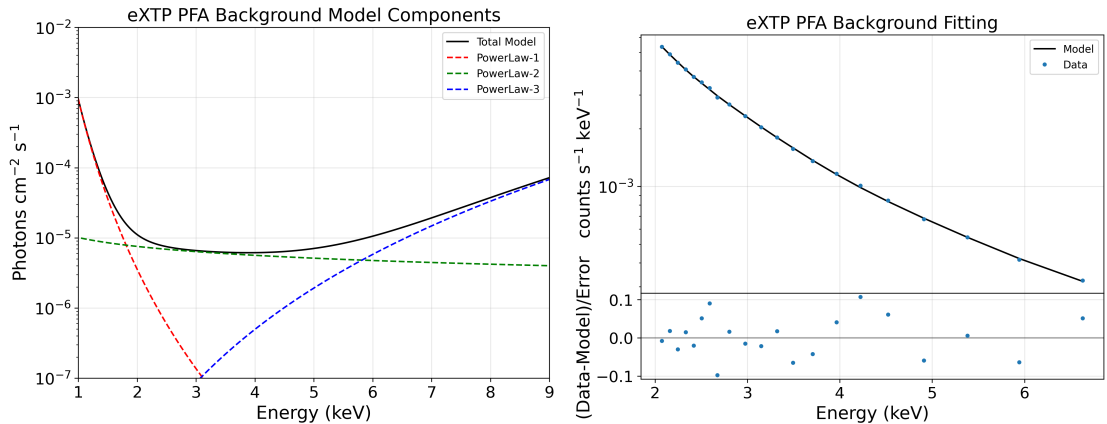

**Figure 2.** The instrumental background model of eXTP-PFA. Left panel: the phenomenological model consisting of three power-law components. The red, green, and blue lines represent the three power-law components, and the black line represents the sum of the three components. Right panel: the fitting result of the GEANT4-simulated background spectrum using the three-component model. The blue data points show the GEANT4-simulated background spectrum, and the black line shows the fitting result.

**Table 1**
Fitting Results of the eXTP-PFA Instrumental Background

| Model | Parameter | Value |
| --- | --- | --- |
| POW1 | $\Gamma$ | 8.0 |
|  | Norm ($10^{-4}$ Photon keV$^{-1}$ cm$^{-2}$ s$^{-1}$) | 9.4 |
| POW2 | $\Gamma$ | 0.4 |
|  | Norm ($10^{-5}$ Photon keV$^{-1}$ cm$^{-2}$ s$^{-1}$) | 1.0 |
| POW3 | $\Gamma$ | −6.1 |
|  | Norm ($10^{-10}$ Photon keV$^{-1}$ cm$^{-2}$ s$^{-1}$) | 1.1 |

**Note.** The background spectrum is fitted with a three-component power-law model in the 2–8 keV energy range. The norm values are presented in units of photon keV$^{-1}$ cm$^{-2}$ s$^{-1}$.

## 3. Verification of eXTP Simulation

In this section, we validate our eXTP simulation framework by conducting systematic comparisons with corresponding IXPE simulations. We perform comprehensive tests across three fundamental source categories, i.e., the point, periodic, and extended sources, to demonstrate that our simulation methodology is reliable and capable of accurately reproducing the spectral-polarimetric, temporal-polarimetric, and spatially resolved polarization properties.

### 3.1. Spectral-polarimetric Analysis of Point Source

First, we perform an eXTP and IXPE simulation for a point source and conduct the spectral-polarimetric fitting. The flux of the source is set to $2 \times 10^{-8}$ erg cm$^{-2}$ s$^{-1}$ in 2–8 keV (same as the Crab). A power-law spectral model ($\Gamma = 2$) combined with a constant polarization model, PD = 40% and Polarization Angle (PA) = 45°, is used. The exposure is set to 10 ks and xPointSource class from IXPEOBSSIM is used.

Xspec tool in HEASOFT is used to perform the spectral-polarimetric fitting. We extracted the Stokes *I*, *Q*, and *U* spectra from the source region (within a radius of 30″) using XPSELECT and XPBIN tool in IXPEOBSSIM. The *I* spectra were rebinned to ensure a minimum of 50 counts in each channel, and a constant 0.17 keV energy rebinning was applied to the *Q* and *U* Stokes spectra, using ftgrouppha tool in HEASOFT. The rebinned $I$, $Q$, and $U$ spectra were fitted with models POWERLAW+POLCONST. The POLCONST is a constant polarization model in Xspec.

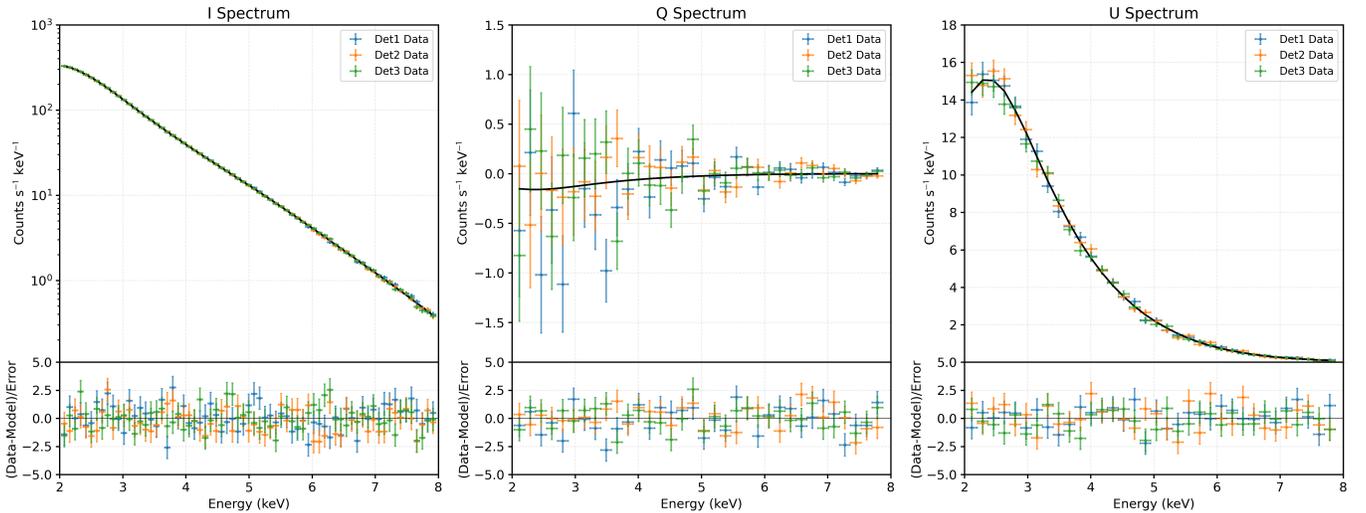

**Figure 3.** Spectropolarimetric fitting results for eXTP point source simulation. The figure shows three panels from left to right: $I$, $Q$, and $U$ Stokes spectra. Each panel contains the spectrum (upper subplot) and residuals (lower subplot). The data points represent simulated observations from three detector units (Det1, Det2, and Det3), while the black solid lines show the best-fit models. The fitting demonstrates excellent agreement between the simulated data and the theoretical model.

**Table 2**
Fitting Results Comparison between IXPE and eXTP Point Source Simulations

| Parameter | Input Value | IXPE Result | eXTP Result |
|---|---|---|---|
| $\Gamma$ | 2.00 | 2.00 ± 0.04 | 2.00 ± 0.02 |
| PD (%) | 40.0 | 40.3 ± 0.4 | 40.2 ± 0.2 |
| PA (°) | 45.0 | 44.5 ± 0.6 | 45.3 ± 0.3 |
| $\chi^2/\mathrm{dof}$ | ⋯ | 1.01 | 1.05 |

The fitting results demonstrate excellent recovery of the input model parameters for eXTP simulations, as shown in Figure 3. Table 2 shows that eXTP achieves significantly improved parameter constraints compared to IXPE, with reduced uncertainties on all fitted parameters. The improvement factor of approximately 2 in measurement precision is owed to the larger effective area, 500 cm$^2$ for eXTP-PFA versus 75 cm$^2$ for IXPE, at 2 keV.

### 3.2. Polarization of Periodic Source

The Crab pulsar serves as an ideal calibration source for X-ray polarimetry due to its well-characterized polarization properties and high flux. We simulate both IXPE and eXTP observations of the Crab pulsar using optical phase-resolved polarization models derived from A. Słowikowska et al. (2009). The pulse profile and phase-resolved spectral properties are based on Chandra observations (M. C. Weisskopf et al. 2011). We employ the xPeriodicPointSource class in IXPEOBSSIM to model the pulsar emission, with a 300 ks exposure.

Figure 4 presents the phase-resolved PD and PA for both IXPE and eXTP simulations. The eXTP simulation results (blue points) exhibit the same factor of 2 improvement in precision as demonstrated in the Section 3.1, with significantly smaller error bars compared to IXPE (orange points). Both IXPE and eXTP simulations successfully reproduce the polarization variations of the model, and eXTP's enhanced sensitivity enables it to more accurately track these theoretical model variations under the same exposure conditions.

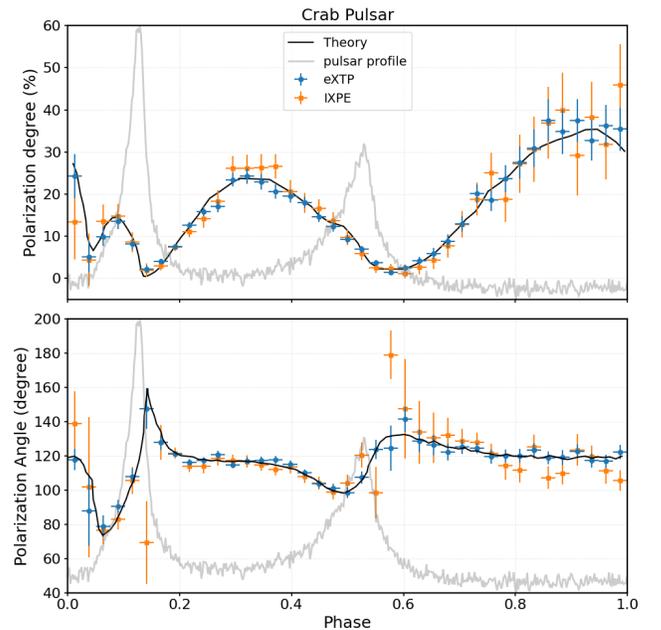

**Figure 4.** Phase-resolved X-ray polarization of the Crab pulsar. Top panel: PD as a function of pulse phase. Bottom panel: PA as a function of pulse phase. The orange points show IXPE simulation results, blue points show eXTP simulation results, and the solid black curve represents the input model. Both simulations used 300 ks exposure time.

### 3.3. Spatially Resolved Polarization Mapping of Extended Source

Spatially resolved polarization mapping is crucial for understanding the complex magnetic field geometries in PWNe. To validate our eXTP simulation framework for extended sources, we perform a comprehensive comparison of spatially resolved polarization mapping between IXPE and eXTP using the well-studied Crab Nebula as a benchmark.

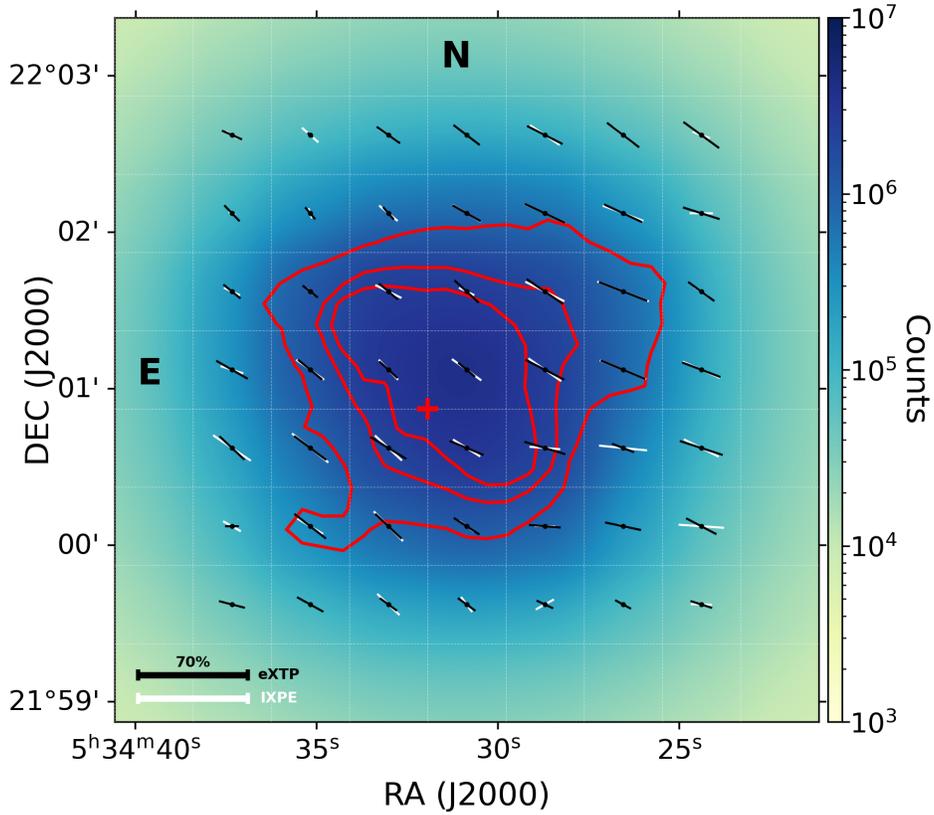

**Figure 5.** Spatially resolved X-ray polarization map of the Crab Nebula. The background color map shows the simulated eXTP X-ray intensity distribution. Red contours represent X-ray intensity levels from Chandra observations. Black vectors show polarization measurements derived from eXTP simulation, while white vectors show those from IXPE simulation. Vector length is proportional to PD, and direction indicates magnetic field orientation (perpendicular to polarization direction). The scale bars indicate 70% polarization for eXTP (black) and IXPE (white).

We simulate the spatially resolved polarization map of the Crab Nebula using morphology derived from Chandra X-ray observations (obsid: 23539). The polarization model is based on spatially resolved X-ray polarization maps obtained from IXPE observations (N. Bucciantini et al. 2023). Both IXPE and eXTP simulations use identical models and a 200 ks exposure.

Figure 5 presents the spatially resolved polarization mapping results with a pixel size of $30'' \times 30''$. The black vectors (eXTP) and white vectors (IXPE) show remarkable agreement in both polarization direction and degree throughout the nebula. Both the IXPE and eXTP simulations show a nearly uniform polarization map. This arises because the input model was derived from existing IXPE observations (which are already convolved with the PSF); the simulation then applies the instrumental response a second time, effectively washing out the detailed toroidal features through a double convolution. This consistency validates the reliability of our simulation framework and confirms that our modifications to IXPEOBSSIM for eXTP-PFA are robust and accurately reproduce the expected polarization properties.

## 4. Simulation Result

In this section, we employ the adjusted IXPEOBSSIM to simulate eXTP-PFA observations of PWNe, showing the enhanced observational potential of eXTP in polarization characteristics of these enigmatic objects.

### 4.1. The X-Ray Polarization Simulation of Faint PWNe

Currently, only a few bright PWNe are observed by IXPE, and only the X-ray polarization of Crab, Vela, MSH 15–52, PWN B0540−69, 3C 58, G21.5, and Kes 75 has been observed. For fainter PWNe, limited by its effective area, it would be quite challenging to obtain a robust observation using IXPE within a reasonable exposure time. eXTP, with its larger effective area, can help us greatly expand the X-ray polarization sample of PWNe. In this subsection, we will simulate the eXTP observation of the faint PWNe and compare their $MDP_{99}$ with that of IXPE.

We selected 12 PWNe from the SNRcat catalog, a comprehensive online database maintained by the Harvard–Smithsonian Center for Astrophysics that provides detailed information about SNRs and their associated PWNe, including their positions, sizes, fluxes, and other physical properties.[4] These 12 sources are ranked based on their nebular flux level from high to low in 2–8 keV energy range: N157B (Y. Chen et al. 2006), G54.1+0.3 (F. J. Lu et al. 2002), Mouse (F. Yusef-Zadeh & B. Gaensler 2005), G189.1−3.0 (C. M. Olbert et al. 2001), G11.18−0.35 (V. M. Kaspi et al. 2001), G327.1−1.1 (M. Sun et al. 1999), G16.7+0.1 (D. J. Helfand et al. 2003), Boomerang (J. P. Halpern et al. 2001a), G0.9+0.1 (B. M. Gaensler et al. 2001), MSH 11−54 (J. P. Hughes et al. 2001), G39.2−0.3 (C. M. Olbert et al. 2003), and Geminga (P. A. Caraveo et al. 2003).

---
[4] https://snrcat.cfa.harvard.edu/

The Chandra ACIS events files of these sources, which have been processed through the standard Chandra data reduction pipeline (chandra_repro), are used as input morphological model, and the xChandraROIModel class of IXPEOBSSIM is used in the simulations (xChandraROIModel can convert the Chandra ACIS event list to a model for IXPE simulation). Since our primary focus is on the $MDP_{99}$, the same and simple polarization model is used for all sources, with a PD of 10% and a PA of 45°. The exposure is set to 1 Ms.

For polarization analysis of faint sources, background subtraction is crucial to ensure accurate results. In our simulations, we have considered two types of backgrounds: the instrumental background and the ambient radiation background surrounding the PWNe. The instrumental background of eXTP is simulated by using xTemplateInstrumentalBkg class from IXPEOBSSIM, with the background template based on GEANT4-simulated data. For IXPE simulations, the instrumental background is directly generated using the xPowerLawInstrumentalBkg class.

The radiation background is handled by selecting a region near the source and subtracted through PCUBE. For the PWNe strongly affected by their SNR background, especially for MSH 11−54 and G11.18−0.35, the background regions are close to the source. For the Mouse, which is close to the nearby bright low-mass X-ray binary SWIFT J1747−3001 $\sim 3'$ away, the background is defined in a region where the distance from the binary is the same as that of the Mouse. The background region shares the same size as the source region in all simulations, and their sizes are determined by the degree of source extension.

The Chandra images and the simulated eXTP images are shown in Figure 6, and the source and background regions are marked with white and yellow circles. The background-subtracted polarization results are summarized in Table 3. The results show that eXTP achieves significantly better $MDP_{99}$ values compared to IXPE. Additionally, according to spectral fitting results and pileup effect reported in B. M. Gaensler et al. (2004), the flux of rebuild photon spectra would be 23% higher than the observed, and the $MDP_{99}$ of Mouse has been corrected based on this ratio.

### 4.2. The Spatially Resolved X-Ray Polarization Map of Extended PWNe

Spatially resolved X-ray polarization is essential for PWNe, which helps us reveal the geometry of magnetic fields and the degree of magnetic field turbulence. Although eXTP has a similar spatial resolution to IXPE, its significantly enhanced effective area will markedly improve its capabilities for observing extended sources. In this subsection, we will show this improvement by simulating the polarization maps for Vela PWN.

Same as above, the Chandra events file of Vela PWN (obsid: 12074) is used as input morphological models, and the xChandraROIModel class is adopted in the simulations. The polarization maps obtained from IXPE observations (IXPE obsid: 01001299) are accepted as the input polarization models. The exposure is set to 1 Ms. For comparison, we also present the simulated IXPE polarization map using the same polarization model and exposure time. The simulation results are presented in Figure 7.

Each pixel with a PD higher than the $MDP_{99}$ is marked in the figure, with the length of the vector representing the PD and the orientation representing the direction of the magnetic field. Compared with IXPE, it is clear that the area with significant polarization detection obtained by eXTP is almost twice as large as that of IXPE.

## 5. Discussion

### 5.1. Detection Capability of eXTP on PWNe

Due to its enhanced capabilities, eXTP offers substantially improved MDP values compared to IXPE across our entire sample, as well as a factor of approximately 2 improvement in PD uncertainty (as established in Section 3.1). Based on the results presented in Table 3, we categorize the 12 PWNe into three distinct groups according to their detectability, using $MDP_{99}$ as the primary indicator.

The most promising targets for eXTP include N157B, Mouse, and G54.1+0.3, which exhibit the lowest MDP values. For these sources, even relatively modest intrinsic PDs of 10%–15% would result in high-significance detections, providing excellent opportunities to obtain robust polarization measurements. For nebula N157B and G54.1+0.3, eXTP would probably enable spatially resolved polarization studies. Moderately detectable targets comprise G189.1+3.0, G11.18-0.35, G16.7+0.1, and G327.1-1.1. To achieve robust polarization measurements within 1 Ms exposure, these sources would require somewhat higher intrinsic PDs of approximately 15%–20%. Such values remain well within the typical range observed in PWNe, thus these targets still present valuable opportunities for expanding the X-ray polarization sample of PWNe. The most challenging targets include MSH 11–54, G39.2-0.3, G0.9+0.1, Boomerang, and especially Geminga. These sources would require remarkably high intrinsic PDs (>20%–30%) for significant detections. However, such observations might still be feasible with larger exposure or if these objects have unexpectedly high polarization.

Beyond expanding the X-ray polarization sample, the enhanced capabilities of eXTP will enable much more comprehensive polarization mapping studies, particularly for moderately extended sources like the Vela PWN. Our simulations demonstrate that eXTP-PFA can achieve effective polarization mapping over an area almost twice as large as that obtained by IXPE, as shown in Section 4.2. This enhanced mapping capability will allow for more detailed studies of magnetic field geometry, providing deeper insights into the complex physics of PWN systems.

### 5.2. Astrophysical Implications of X-Ray Polarization of PWNe

PWNe at different evolutionary stages exhibit distinct X-ray polarization characteristics, each offering unique insights into PWN physics. Based on their evolutionary phases, the PWNe in our sample can be classified into three main groups.

The first group represents young PWNe similar to the Crab, including MSH 11−54, G11.18−0.35, G54.1+0.3, and G0.9 +0.1. MHD simulations show that PWNe at this stage typically exhibit a jet-torus structure with toroidal magnetic fields (S. V. Bogovalov & D. V. Khangoulian 2002; Y. E. Lyubarsky 2002; S. S. Komissarov & Y. E. Lyubarsky 2003), where the jet direction typically aligns with the integrated polarization direction (L. Del Zanna et al. 2004, 2006; N. Bucciantini et al. 2005; D. Volpi et al. 2009), as confirmed by IXPE observations of the Crab (N. Bucciantini et al. 2023). Therefore, for those obscure

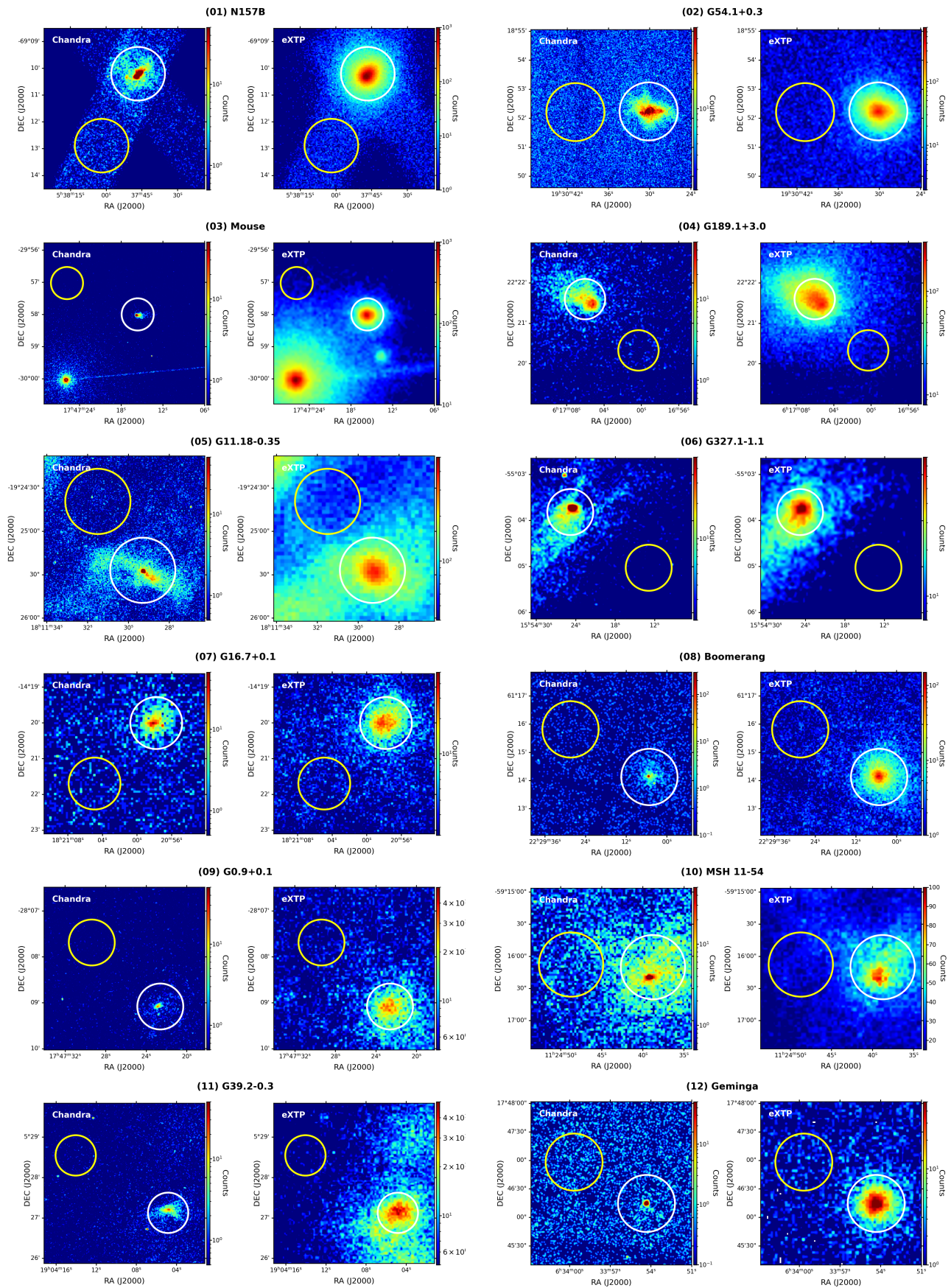

**Figure 6.** Comparison of Chandra X-ray images (left panels) and simulated eXTP images (right panels) for 12 PWNe. From top to bottom and left to right: N157B, G54.1+0.3, Mouse, G189.1-3.0, G11.18-0.35, G327.1-1.1, G16.7+0.1, Boomerang, G0.9+0.1, MSH 11–54, G39.2-0.3, and Geminga. The white circles indicate the source regions used for polarization analysis, while the yellow circles show the background regions.

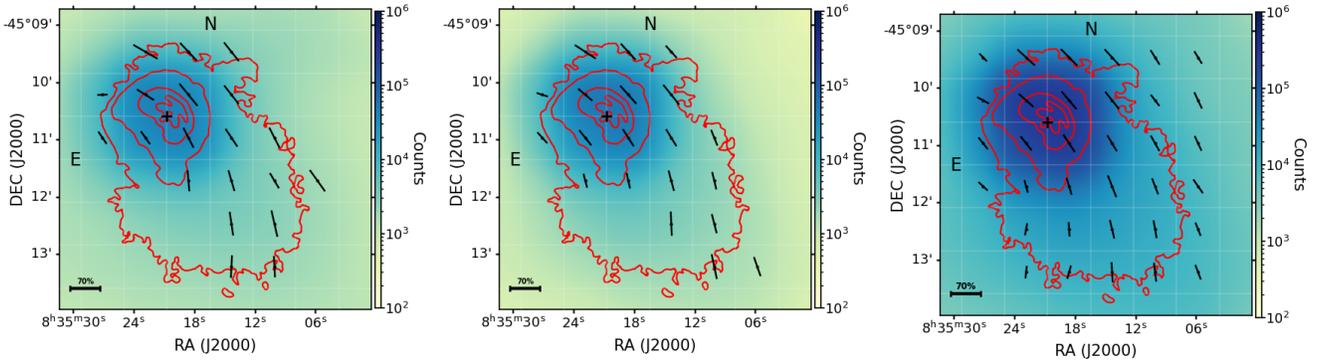

**Figure 7.** Comparison of X-ray polarization maps for Vela PWN. Left panel: IXPE observational data. Middle panel: simulated IXPE polarization map. Right panel: simulated eXTP-PFA polarization map. The black vectors indicate the polarization direction and degree, with only pixels having PDs higher than MDP$_{99}$ shown. The red contours represent X-ray intensity levels obtained from the Chandra observation.

**Table 3**
MDP$_{99}$ Values for the 12 PWNe

| No. | Source Name | R.A. (J2000) | Decl. (J2000) | ObsID | IXPE MDP (%) | eXTP MDP (%) |
|---|---|---|---|---|---|---|
| 01 | N157B | 05$^h$37$^m$47$^s$.6 | −69°10′ 20″.0 | 2783 | 8.4 | 4.3 |
| 02 | G54.1+0.3 | 19$^h$30$^m$30$^s$.2 | +18°52′ 14″.2 | 9109 | 8.9 | 4.6 |
| 03 | Mouse | 17$^h$47$^m$15$^s$.9 | −29°58′ 02″.0 | 2834 | 8.2 | 4.8 |
| 04 | G189.1−3.0 | 06$^h$17$^m$05$^s$.0 | +22°21′ 27″.0 | 13736 | 14.3 | 7.7 |
| 05 | G11.18−0.35 | 18$^h$11$^m$29$^s$.4 | −19°25′ 25″.0 | 14831 | 16.6 | 9.3 |
| 06 | G327.1−1.1 | 15$^h$54$^m$31$^s$.0 | −55°05′ 40″.0 | 13768 | 21.8 | 12.1 |
| 07 | G16.7+0.1 | 18$^h$20$^m$57$^s$.8 | −14°20′ 09″.6 | 3845 | 24.6 | 12.7 |
| 08 | Boomerang | 22$^h$29$^m$05$^s$.2 | +61°14′ 08″.9 | 2787 | 23.5 | 13.5 |
| 09 | G0.9+0.1 | 17$^h$47$^m$23$^s$.6 | −28°09′ 35″.0 | 1036 | 26.3 | 14.7 |
| 10 | MSH 11−54 | 11$^h$24$^m$39$^s$.5 | −59°15′ 56″.4 | 12555 | 27.4 | 15.3 |
| 11 | G39.2−0.3 | 19$^h$04$^m$06$^s$.0 | +05°27′ 00″.0 | 1988 | 27.8 | 15.9 |
| 12 | Geminga | 06$^h$33$^m$54$^s$.3 | +17°46′ 14″.7 | 14692 | 50.0 | 29.7 |

PWNe, X-ray polarization observations can thus serve as powerful probes of their morphological structures.

MSH 11–54, with an age comparable to the Crab (∼1600 yr), displays a compact core surrounded by diffuse extended emission (J. P. Hughes et al. 2001). J. P. Hughes et al. (2003) identified significant residual structures along the northwest–southeast direction after modeling the diffuse emission with Gaussian profiles, interpreting these features as potential jet structures. Given the theoretical prediction that integrated PA should align with jet direction, X-ray polarimetry of MSH 11–54 could definitively confirm whether these residual features represent genuine jet structures.

G11.18-0.35 exhibits prominent outflow structures without a significant torus component, suggesting a jet-dominated configuration (V. M. Kaspi et al. 2001; M. S. E. Roberts et al. 2001). Remarkably, despite the absence of visible jet-like structures at radio wavelengths, radio polarization observations reveal magnetic field orientations aligned with the jet direction (Y. Zhang et al. 2025), providing indirect evidence for jet structures. X-ray polarimetry could provide direct confirmation of its jet-dominated nature by measuring the integrated polarization direction.

The second group comprises intermediate-age PWNe including N157B, G327.1-1.1, G16.7+0.1, and Boomerang. These systems are believed to be interacting with reverse shocks from their parent supernova explosions, resulting in more complex morphological and polarization characteristics.

The reverse shock strips away outer nebular material and exposes fresh nebular material. Thus, the system could exhibit higher PD and improved consistency between X-ray and radio PD measurements, as observed in Vela PWN and MSH 15-52 (F. Xie et al. 2022; K. Liu et al. 2023; R. W. Romani et al. 2023). The X-ray polarization studies of these PWNe could provide crucial insights into the interaction processes between PWNe and reverse shocks, helping us understand the physics behind the variety of their morphologies.

G327.1-1.1 exhibits a snail-like morphology with an unclear origin, particularly its arcminute-scale double prong-like structures extending northwest (M. Sun et al. 1999; Y. K. Ma et al. 2016). T. Temim et al. (2015) suggested the entire structure resembles a funnel-shaped outflow, with the prongs possibly representing limb-brightened edges rather than separate structures. A spatially resolved X-ray polarization observation could help understand their physical nature.

Boomerang, another typical second-stage PWN, shows high polarization in radio wavelengths, revealing magnetic field structures highly consistent with those of the Vela PWN in radio bands (J. P. Halpern et al. 2001a, 2001b; R. Kothes et al. 2001; R. Kothes et al. 2006; Y. Fujita et al. 2021). Due to its low brightness, eXTP is expected to achieve an MDP of 13.6% in a 1 Ms observation. However, considering its similarity to Vela, its X-ray PD might exceed 40%, potentially yielding significant polarization results with eXTP observations.

The third group comprises mature PWNe including G189.1+3.0, Mouse, and Geminga, which are characterized by bow-shock structures formed as pulsars move through their environment. G189.1+3.0 represents a PWN interacting with the SNR edge (F. Bocchino & A. M. Bykov 2001; C. M. Olbert et al. 2001), while Mouse and Geminga exhibit bow shocks formed as their host pulsars move through the ISM (B. M. Gaensler et al. 2004; F. Yusef-Zadeh & B. Gaensler 2005; N. Klingler et al. 2018). To date, no bow-shock PWN has been detected in X-ray polarization, representing a significant observational gap in our understanding of these evolved systems. These mature PWNe offer unique opportunities for studying termination shock physics. Without the obscuration from extended nebular material, direct observation of X-ray emission near the termination shock becomes feasible. This emission preserves crucial information about magnetic field structures in particle acceleration regions, making them particularly valuable for constraining fundamental acceleration mechanisms.

## 6. Conclusion

In this paper, we have demonstrated the enhanced X-ray polarization capabilities of eXTP-PFA for studying PWNe. We present the first implementation of using modified IXPEOBSSIM to simulate eXTP-PFA X-ray polarization observations, establishing a robust methodology that can be readily extended to high-energy astrophysical targets including black hole X-ray binaries, active galactic nuclei, gamma-ray bursts, and so on.

Our methodology has been comprehensively validated through detailed comparisons. The validation encompasses phase-resolved polarization studies, spectropolarimetric analysis with Xspec, and spatially resolved polarization mapping, all demonstrating excellent agreement between IXPE and eXTP simulation.

We evaluated the polarimetric performance of eXTP-PFA for PWNe with readapted IXPEOBSSIM. Our simulations show that eXTP-PFA provides a factor of approximately 2 improvement in $MDP_{99}$ values compared to IXPE across all targets. The brightest PWNe in our sample (N157B, G54.1+0.3, and Mouse) reach impressive $MDP_{99}$ values of 4%–5%. For moderately challenging targets like G189.1+3.0 and G11.18-0.35, significant detections remain feasible with the typical intrinsic PDs of ∼15%–20%. For the most challenging sources such as MSH 11–54, G39.2-0.3, and Geminga, they would need a high PD (>20%–30%) to achieve significant detections.

The spatial mapping capabilities of eXTP-PFA are particularly notable, as demonstrated by our Vela PWN simulations. With equivalent exposure times, eXTP-PFA can effectively map polarization over an area nearly twice as large as IXPE, enabling more comprehensive studies of magnetic field geometries and their spatial variations in extended sources.

These simulation results show that eXTP-PFA will significantly expand the sample of PWNe with robust X-ray polarization measurements beyond the few bright sources currently accessible to IXPE, which would shed light on the relativistic particle acceleration, energy conversion processes, and complex plasma dynamics in PWNe.

## Acknowledgments

This work is supported by National Natural Science Foundation of China (grant Nos. 12422306 and 12373041), and Natural Science Foundation of Guangxi (grant Nos. 2025GXNSFDA02850001), and Bagui Scholars Program (XF). This work is also supported by the Guangxi Talent Program ("Highland of Innovation Talents"). K.L. is supported by Innovation Project of Guangxi Graduate Education (grant No. YCBZ2025065). This paper employs a list of Chandra datasets, obtained by the Chandra X-ray Observatory, contained in doi:10.25574/cdc.530.

## ORCID iDs

Kuan Liu https://orcid.org/0009-0007-8686-9012
Fei Xie https://orcid.org/0000-0002-0105-5826
Ming-Yu Ge https://orcid.org/0000-0002-3776-4536
Wei Deng https://orcid.org/0000-0002-9370-4079
En-Wei Liang https://orcid.org/0000-0002-7044-733X